# Applications of aligned nanofiber for tissue engineering

Gayatri Patel[a], Louis-S. Bouchard[b*]

[a]Department of Bioenginnering, UCLA, USA

[b]Department of Chemistry and Biochemistry, UCLA, USA

Corresponding author:
* E. Mail: louis.bouchard@gmail.com (Louis Bouchard)

**Abstract**

In tissue engineering, we seek to address comprehensive tissue repair and regeneration needs. Aligned nanofibers have emerged as powerful and versatile tools, attributable to their structural and biochemical congruence with the natural extracellular matrix (ECM). This review delineates the contemporary applications of aligned nanofibers in tissue engineering, spotlighting their implementation across musculoskeletal, neural, and cardiovascular tissue domains. The influence of fiber alignment on critical cellular behaviors—cell adhesion, migration, orientation, and differentiation—is reviewed. We also discuss how nanofibers are improved by adding growth factors, peptides, and drugs to help tissues regenerate better. Comprehensive analyses of in vivo trials and clinical studies corroborate the efficacy and safety of these fibers in tissue engineering applications. The review culminates with exploring extant challenges, concurrently charting prospective avenues in aligned nanofiber-centric tissue engineering.

**Contents**

## 1. Introduction

Tissue engineering aims to heal or replace damaged tissues using cells, scaffold materials, and bioactive molecules [1]. A key to its success is the development of scaffolds that mimic the natural extracellular matrix (ECM) to guide cell behavior and tissue growth [1]. Aligned nanofibers have become notable among scaffold materials due to their unique structural and functional properties [1].

Their nanoscale dimensions and structure mirror the native ECM [2], making them effective for guiding cell growth and differentiation [3, 4]. Their alignment resembles patterns found in various tissues, such as muscles and nerves, facilitating proper cell orientation and function [5]. Moreover, their design, which boasts a high surface-to-volume ratio, promotes efficient nutrient exchange and cell signaling [6].

Multiple methods exist to produce these nanofibers, including electrospinning [7], self-assembly [8], and template-assisted techniques [9]. Electrospinning, in particular, stands out for its versatility and simplicity, offering control over nanofibers' alignment, diameter, and mechanical properties [10].

Aligned nanofibers have shown potential in diverse applications. For example, in musculoskeletal tissue engineering, they assist in regenerating skeletal muscle, bone, cartilage, and tendons [11]. They've also been used in neural tissue engineering to guide axonal growth and support neural cell differentiation [12], and in cardiovascular applications to promote blood vessel organization and growth [13, 14]. Cells adhere, move, and orient more effectively on these fibers, mirroring behaviors observed in natural tissues.

Incorporating bioactive molecules, such as growth factors, peptides, or drugs, enhances their therapeutic potential [15, 16]. These nanofibers can methodically release these agents, directing cells to proliferate and differentiate [17]. In this review, we provide a comprehensive analysis of the applications of aligned nanofibers in tissue engineering, as outlined in Scheme 1.

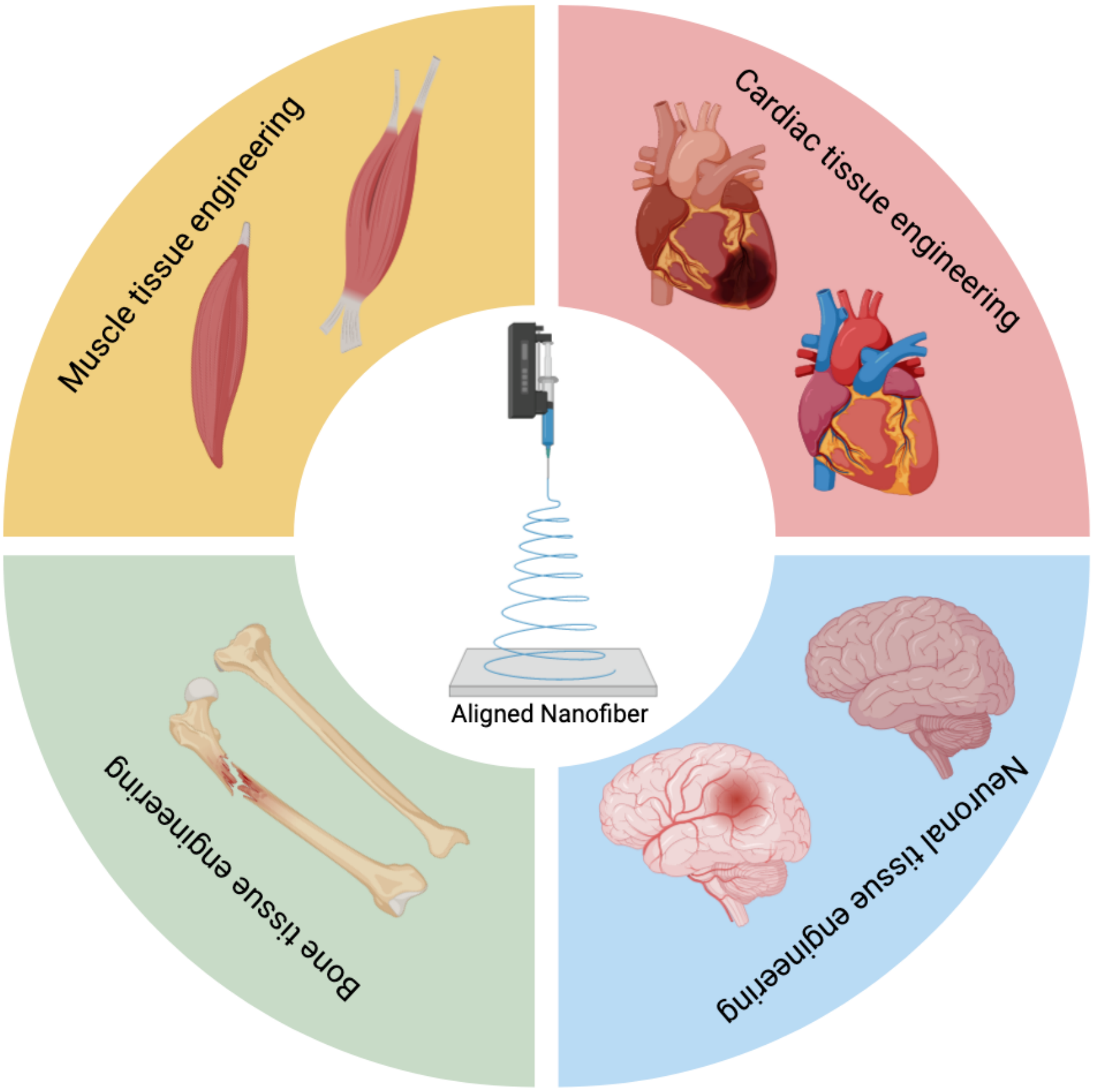

**Scheme 1.** Applications of aligned nanofiber for tissue engineering. Created by Biorender.com.

## 2. Electrospinning aligned nanofiber

Electrospinning is a versatile and popular technique for producing nanofibers from various polymer solutions [18]. Aligned nanofibers refer to fibers oriented in a specific direction, offering unique mechanical, electrical, and structural properties. The alignment of nanofibers can be controlled during the electrospinning process through several methods, such as electrospinning setup modification, collector design, and coaxial electrospinning [18]. Aligned nanofibers have numerous applications, including tissue engineering, drug delivery [19], filtration membranes [20], sensors [21], and flexible electronics [22]. They can offer improved mechanical properties, enhanced anisotropy, and specific functionalities compared to randomly oriented nanofibers. Recent aligned nanofiber studies for tissue regeneration are summarized in **Table 1**. Here, we have discussed the tissue regeneration application of aligned nanofiber, especially bone, neuron, cardiac, and muscle tissue.

**Table1**. Aligned nanofiber for different tissue engineering.

| Material | Cell Type | Target tissue | Biological outcome | Ref |
|---|---|---|---|---|
| PLGA/ oHA-Collagen/HAP | PIECs, MC3T3-E1 and Schwann cells | Bone | -Orderly cell growth and migration<br>-Induced vascularization with the addition of oHA-Col/HAP<br>-Stimulated collagen deposition in vivo | [23] |
| PLGA | UMR106 and MC3T3-E1 | Bone | -Filopodia formation in osteoblast cells leads to cell growth and proliferation | [24] |
| PLLA | hMSCs | Bone | -Cell migration increased ten times in the parallel direction as opposed to the perpendicular direction<br>-Polydopamine coating improved recovery rate in vivo | [25] |
| PLLA | MSCs | Bone | -Stem cell differentiation into tendon tissue<br>- Enhanced osteogenic markers (Bmp4, Ocn, Runx2) upon culture on randomly aligned nanofiber<br>- | [26] |
| PLA | Pre-osteoblast cells | Bone | - PLA/nHA 20% presented desirable cell biocompatibility to aligned nanofiber | [27] |
| PLA | MC3T3-E1 and CRL-2593 | Bone | -Improved cell proliferation and attachment compared to bare PLA<br>- Excellent bone healing capabilities in vivo | [28] |
| PLA/MWCNTs | Osteoblast | Bone | -Electrical current of 100uA enhanced cells growth on aligned fiber | [29] |
| PLA/ Gelatin | MG-63 and fibroblast L929 | Bone | -Multiple pore sizes in membrane allowed proper nutrient supply<br>-prevention from bacterial infection and triggering osteoinductive signals | [30] |
| SF/PCL | BMSCs | Bone | -Radially aligned scaffold improved rate of bone regeneration in vivo | [31] |
| PCL | HUVECs and MSC | Bone | -Good adherence and proliferation on nanofiber scaffold | [32] |
| PCL | PC12 and hADSCs | Neuron | -Cell viability, differentiation, and neurite length enhanced with the lignin content in vitro<br>-15% lignin content showed the highest nerve regeneration in vivo | 33 |
| PCL | DRGs | Neuron | - Neurite outgrowth enhanced up to~2.2-fold with the presence of micro-RNA<br>-Micro RNA led to a longer length of nerve regeneration in the sciatic injury model | 34 |
| PCL/SF/ CNT | PC12 and DRGs | Neuron | -Induced neurite alignment and extension | 35 |
| PLGA | Primary neurons | Neuron | -The aligned nanofiber promoted neuron viability and neurite expression in the presence of LysoGM1 in vitro<br>- Significantly improved the neurogenesis and minimized the glial scar formation in vivo | 36 |
| PLGA | Neural and Schwann cells | Neuron | -Increased cell adhesion<br>- Accelerated nerve regeneration in the presence of MAP and IKVAV | 37 |
| PLLA/PCL | Rat bone marrow-derived macrophages and Schwan cells | Neuron | - Aligned nanofiber significantly promoted macrophage elongation and polarization towards a pro-healing phenotype<br>- Enhanced Schwann cells proliferation and migration *in vitro*<br>- Exhibited Schwann cells infiltration and axonal regeneration *in vivo* | 38 |
| PLGA/MWCNT | PC12, and DRGs | Neuron | -Cell proliferation and differentiation increased with the content of MWCNT<br>- Neurite length, cell proliferation, differentiation, and myelination enhanced on the aligned nanofiber with electrical stimulation | 39 |
| PLLA | Human SH-SY5Y neuroblastoma cells | Neuron | -Aligned nanofiber improved neurite outgrowth by 30% | 40 |
| PLLA | DRGs | Neuron | - Neurite adhesion and extension significantly improved in aligned fiber scaffold | 41 |
| P(3HO)/P(3HB) | NG108-15 neuronal and RN22 Schwann cells | Neuron | - Increased neuronal growth and differentiation in Schwann cells | 42 |
| PPy/CAB | Human neuroblastoma cells | Neuron | - Enhanced cell adhesion and neurite outgrowth under electrical stimulation | 43 |
| SF/Melanin | Human neuroblastoma cells | Neuron | - Aligned nanofiber promoted cell alignment and the neuronal differentiation | 44 |
| PLGA | Human induced pluripotent stem cells | Cardiac | -Cells seeded on aligned fiber obtained cardiac tissue-like constructs (CTLCs)<br>- CTLCs improved significant cardiomyocyte survival, upregulated cardiac biomarkers and cardiac function in vivo | 14 |
| PCL/SF/CNT | Cardiomyocytes | Cardiac | -Aligned nanofiber controlled cellular orientation and promoted cardiomyocyte maturation<br>-Accelerated cardiac tissue regeneration and upregulated cardiac biomarkers | 45 |
| PCL | HL-1 cardiomyocytes and human induced pluripotent stem cell-derived cardiomyocytes | Cardiac | -Showed cell alignment and cell orientation<br>- The scaffold showed functional multiaxial contraction | 46 |
| PCL | Induced pluripotent stem cell-derived cardiac progenitor cells | Cardiac | - Improved functional cardiomyocytes differentiation<br>-Significantly increased the cardiac marker gene<br>- The aligned scaffolds displayed spontaneous synchronized intracellular $Ca^{2+}$ oscillation and cell contraction | 47 |
| PCL | Human-induced pluripotent stem | Cardiac | -Cell morphology changed to rod like shape<br>-Upregulated the cardiomyocytes mature gene (MYH7) | 48 |
| | cell -derived cardiomyocytes | | | |
| PLA/PEG/Collagen | H9C2 cardiomyocyte cell | Cardiac | -Showed cell alignment and cardiomyocytes morphology | 49 |
| PLA/Chitosan | Cardiomyocytes | Cardiac | -Enhanced cell viability, cell elongation and production of contractile protein, and increased cardiomyocyte differentiation | 50 |
| PU/EC | H9C2 rat cardiac myoblasts | Cardiac | -Significantly improved the cell viability, guidance, and proliferation | 51 |
| PEUU/Gelatin | Cardiomyocytes | Cardiac | -Cell proliferation increased with a 70:30 PEUU/gelatin ratio | 52 |
| PCL/Collagen | Myoblasts/ ADSC/ Schwann cells | Muscle | -Myogenic gene was upregulated in the coculture condition of the three cell lines<br>-Myogenic differentiation enhanced with the presence of Schwann cells | 53 |
| PCL/ D-ECM | - | Muscle | - Significantly upregulated the myogenic markers such as MyoD and myogenin<br>- Accelerated muscle regeneration in a Volumetric muscle loss murine model | 54 |
| PCL/Collagen-I | Myoblasts and mesenchymal stem cells | Muscle | - Myogenic differentiation increased in coculture under hepatocyte growth factor and insulin-like growth factor-1 treatment | 55 |
| PCL-Collagen I | Myoblasts, BMSC and ADSC | Muscle | - Upregulated late myogenic markers such as Myosin heavy chain 2, ACTN2, and MYOG in the coculture condition<br>-Cell viability increased in ADSC coculture condition | 56 |
| P(LLA-CL)/ Collagen | Placental stem cells | Muscle | -Enhanced myoblast differentiation by upregulating myoblast-specific markers such as α-SMA, desmin, and collagen type 1, 3 | 57 |
| PCL/SF /PANI | C2C12 myoblasts | Muscle | -Induced cellular alignment and elongation | 58 |
| | | | -Myogenic differentiation increased by upregulating MHC protein | |
| PCL | Murine skeletal muscle cells | Muscle | - Enhanced adhesion, proliferation, and differentiation of muscle cells<br>- myogenic gene expression increased with the PEDOT:PSS coating | 59 |
| PCL | H9c2 myoblast cells | Muscle | -Guided myoblasts alignment and enhanced myoblasts differentiation into long and thick myotubes<br>-Myotube formation accelerated with the gold coating on nanofiber surfaces | 60 |
| PCL | C2C12 myoblasts | Muscle | -Significantly increased myoblast proliferation, alignment, and facilitated the formation of myotubes<br>- Upregulated myogenic genes such as MyoD, myogenin, and troponin T | 61 |
| Alginate/PEO | C2C12 myoblast cells | Muscle | -- Facilitated myoblast adhesion and alignment<br>- Cell orientation and myotube formation significantly better in aligned nanofiber | 62 |
| PCL/PLLA | Wharton's Jelly-derived mesenchymal stem cell | Muscle | - Induced differentiation of WJ-derived UC-MSCs into smooth muscle cells (SMC)<br>-Enhanced SMC markers with the sustained release of TGF-β1 | 63 |
| P(EDS)UU-POSS | C2C12 mouse myoblast cell | Muscle | -The aligned nanofiber guided myoblast alignment, enhanced sarcomere formation, and promoted myotube fusion and myofiber maturation | 64 |

**Abbreviations:** PLGA: poly (lactic-co-glycolide); PLLA: Poly-L-lactic Acid; PLA: Polylactic acid; MWCNT: Multiwalled Carbon nanotube; SF: Silk Fibroin; PCL: Polycaprolactone; P(3HO) Poly-3-hydroxyoctanoate; P(3HB): Poly(3-Hydroxybutyrate); PPy: Polypyrrole; CAB: cellulose acetate butyrate; PEG: Polyethylene glycol; PU: polyurethane; EC: Ethyl cellulose; PEUU: Polyester urethane urea; PANI: Polyaniline ; PEO: Polyethylene oxide; P(EDS)UU-POSS: polyurethane-urea- polyhedral oligomeric silsesquioxane; oHA-Col/HAP:hyaluronic acid oligosaccharide-collagen mineralized microparticles; PIECs: Porcine iliac artery endothelial cells; hMSCs: human mesenchymal stem cells; BMSC: Bone marrow derived stem cells; hADSCs: Human adipose derived stem cells; DRGs: Dorsal root ganglion cells.

### 2.1. Aligned nanofibers for bone tissue engineering

Bone tissue engineering seeks to devise strategies to repair, regenerate, or replace damaged or lost bone tissue. The overarching goal is to craft functional, biocompatible scaffolds and materials that abet bone regeneration following trauma, disease, congenital defects, or other bone loss conditions. Scaffold design is pivotal, aiming to mirror the natural bone environment and bolster the emergence of new tissue. In the realm of bone tissue engineering, aligned nanofibers have emerged as a focal point owing to their potential to replicate the native ECM of bone tissue [23, 25, 65]. These nanofibers, often crafted from a mix of synthetic and natural polymers, effectively mimic the mechanical properties of bone. The benefits of aligned nanofibers include: a. Providing physical cues to guide cells, promoting tissue growth akin to the native bone; b. Enhancing mechanical properties resembling natural bone; c. Guiding blood vessel formation in directions consistent with the bone matrix; d. Achieving localized and sustained release to bolster bone regeneration [23, 66, 67].

In a pivotal study, Liu et al. employed aligned poly(lactic-co-glycolide) (PLGA) nanofibers, infused with hyaluronic acid oligosaccharide-collagen mineralized microparticles (oHA-Col/HAP), to create a bone scaffold. This scaffold was designed to guide cell proliferation. Notably, its aligned structure markedly bolstered strength in the parallel direction, standing in contrast to a randomly oriented scaffold. *In vitro* analyses showcased its ECM-like structural organization, prompting orderly cell growth and migration. Furthermore, the oHA-Col/HAP@PLGA scaffold drove vascularization, as evident from the heightened expression of platelet endothelial cell adhesion molecule (PECAM) and vascular endothelial growth factor (VEGF). *In vivo* analyses vouched for the scaffold's biocompatibility and showcased substantial collagen deposition [23].

Shifting focus to nanofiber-cell interactions, Stachewicz et al. deployed 3D imaging via focused ion beam (FIB)-scanning electron microscopy (SEM). Their exploration of interactions between PLGA and osteoblasts revealed an organization that spurred filopodia formation, thereby driving cell growth. This organized growth starkly contrasted the dispersed cell growth observed on randomly oriented PLGA scaffolds [24].

Lee et al. innovated by modifying aligned fibers' surfaces, crafting polydopamine (PDA)-coated poly(l-lactic acid) (PLLA) fibers to culture hMSCs. Their *in vitro* assessments spotlighted cell adhesion patterns aligning with fiber directionality. More impressively, cell movement surged over tenfold in the parallel direction compared to its perpendicular counterpart. Additionally, an *in vivo* study involving a mouse model underscored the rapid bone healing capabilities of these aligned PDA-coated PLLA fibers [25]. Furthermore, bone morphogenetic protein-2 (BMP-2) was immobilized on PDA-coated PLLA nanofiber. The manner of cell growth was highly dependent on the aligned fiber (Fig 1 A), but there wasn’t much difference in alkaline phosphatase (ALP) activity and calcium mineralization of hMSCs across 14 days of in vitro studies compared to results from randomly oriented fiber. However, *in vivo* evaluation revealed a higher rate of bone regeneration in the random group ($69.2 \pm 0.1\%$) compared to the aligned fiber ($65.7 \pm 0.4\%$) (Fig 1 B). The regenerated tissue used for nanoindentation possessed an anisotropic feature for the model treated with an aligned scaffold. It deposited collagen in a highly ordered fashion to mimic the natural bone structure. The author proposed a possible in vivo collagen matrix assembly mechanism guided by aligned nanofiber (Fig 1 C) [68].

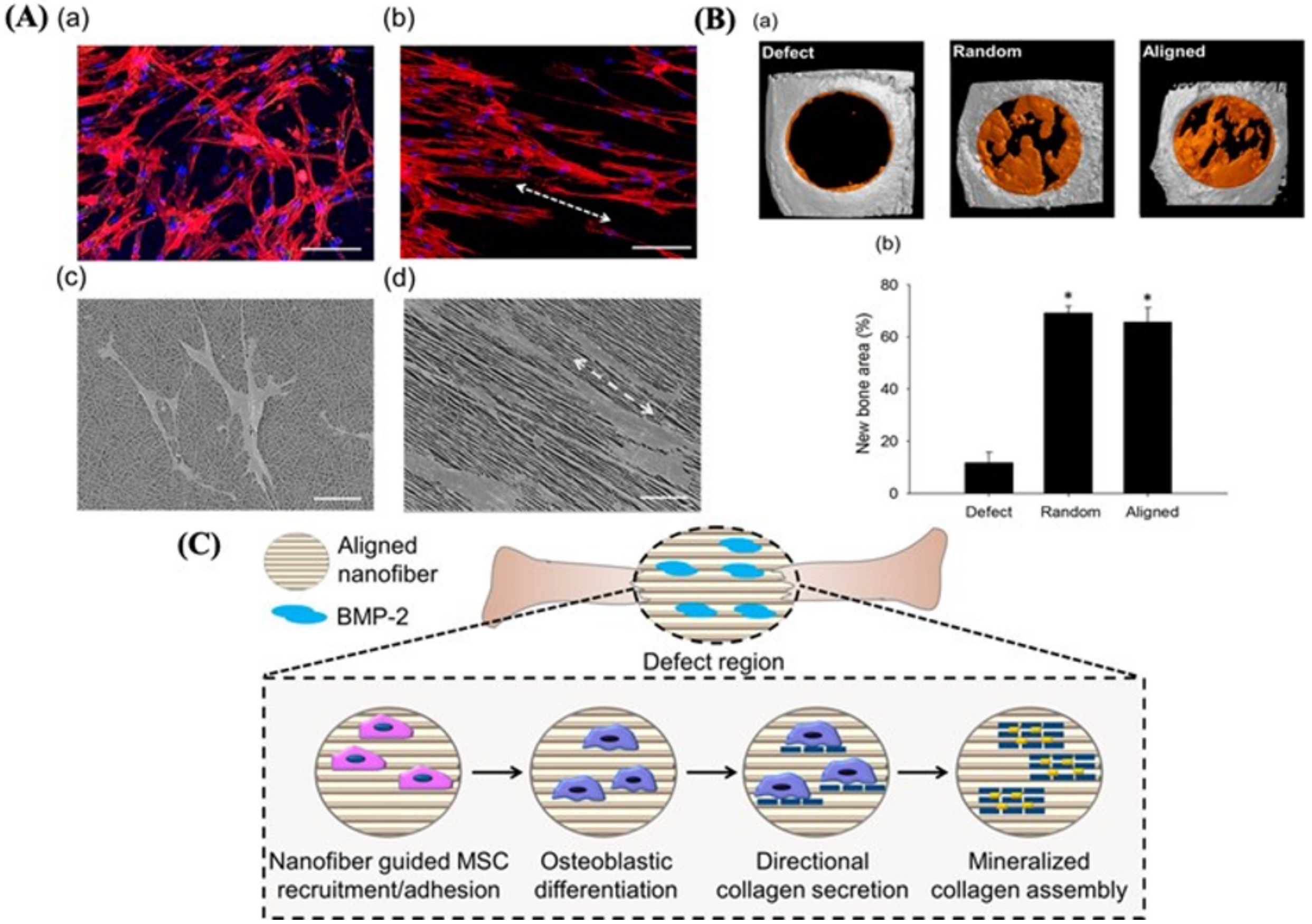

**Fig 1.** (A) Fluorescence images of hMSC adhere to (a) random and (b) aligned nanofibers stained with F-actin (red) and nuclei (blue) after 24 h of culture (scale bar: 100 μm). SEM images of hMSCs on (c) random and (d) aligned nanofibers after 24 h of culture (scale bar: 20 μm). (B) Radiographic analysis of skull bone after two months of post-surgery (a) micro-CT images of skull bone implanted with random and aligned nanofiber (b) the quantitative values of regenerated new bone area from each group. (C) Schematic mechanism of in vivo collagen assembly as instructed by the implanted aligned nanofibers [68].

In another study, Fu et al. endeavored to enhance PLLA nanofibers by integrating calcium silicate nanowires via a hydrothermal process. Analyses of these modified scaffolds—both randomly oriented and aligned—revealed the nanowires' potential in amping up the biocompatibility of PLLA nanofibers [69]. Echoing the importance of biochemical signals, Yin et al.'s study demonstrated that aligned PLLA scaffolds fostered mature tendon tissue formation, whereas randomly oriented ones induced chondrogenesis and ossification [26].

Polylactic acid (PLA), the same family with PLLA, has also significant research attention. Lopresti et al. integrated both micrometric (μ) and nanometric(n) HA particles into aligned PLA electrospun scaffolds. Their *in vitro* studies underscored the biocompatibility of pre-osteoblast cells with nHA-aligned PLA scaffolds, marking the aligned PLA/nHA 20% composition as optimal [27]. However, PLA's limited cell affinity posed challenges. Addressing this, Hwang et al. coated PLA fibers with Tantalum (Ta), which exhibited enhanced osteoconductivity in a rabbit calvarial defect model [28]. Recognizing the potency of electrical cues in guiding cell extension, Shao et al. conceived a conductive PLA/multiwalled carbon nanotubes (MWCNTs) scaffold, highlighting the synergistic impact of topographic and electrical signals on osteoblast growth [29]. Similarly, aiming for enhanced bone regeneration, Shakeri et al. developed a multilayer guided bone regeneration (GBR) approach with PLA-infused simvastatin (SIM) complemented by a gelatin (GT) layer enriched with thymol (THY). Their novel GT/PLA nanofiber showcased potential in bacterial infection prevention and osteoinductive signal release [30].

Electrospinning for bone tissue engineering is a promising avenue, but its 2D nature often results in dense nanofiber mats that hinder cell proliferation. A notable advancement was made by Xiao et al., who fabricated 3D polycaprolactone (PCL) scaffolds using gas foaming combined with biomineralization for bone regeneration in a cranial defect model (Fig 2 A). They applied a silk fibroin coating on the radially and laterally aligned scaffold, enhancing cell arrangement and proliferation (Fig 2 B, C) [31]. For enhanced bone regeneration after injury, vascularization is essential. Recognizing this, researchers cocultured MCS and endothelial cells. Yao et al.'s study with HUVECs and MSCs on PCL electrospun fibers highlighted the significance of the HUVECs in osteogenesis [32]. Reinwald et al. explored how intermittent hydrostatic pressure affects bone marrow-derived MSCs, revealing the impact of scaffold orientation on osteogenic markers [70].

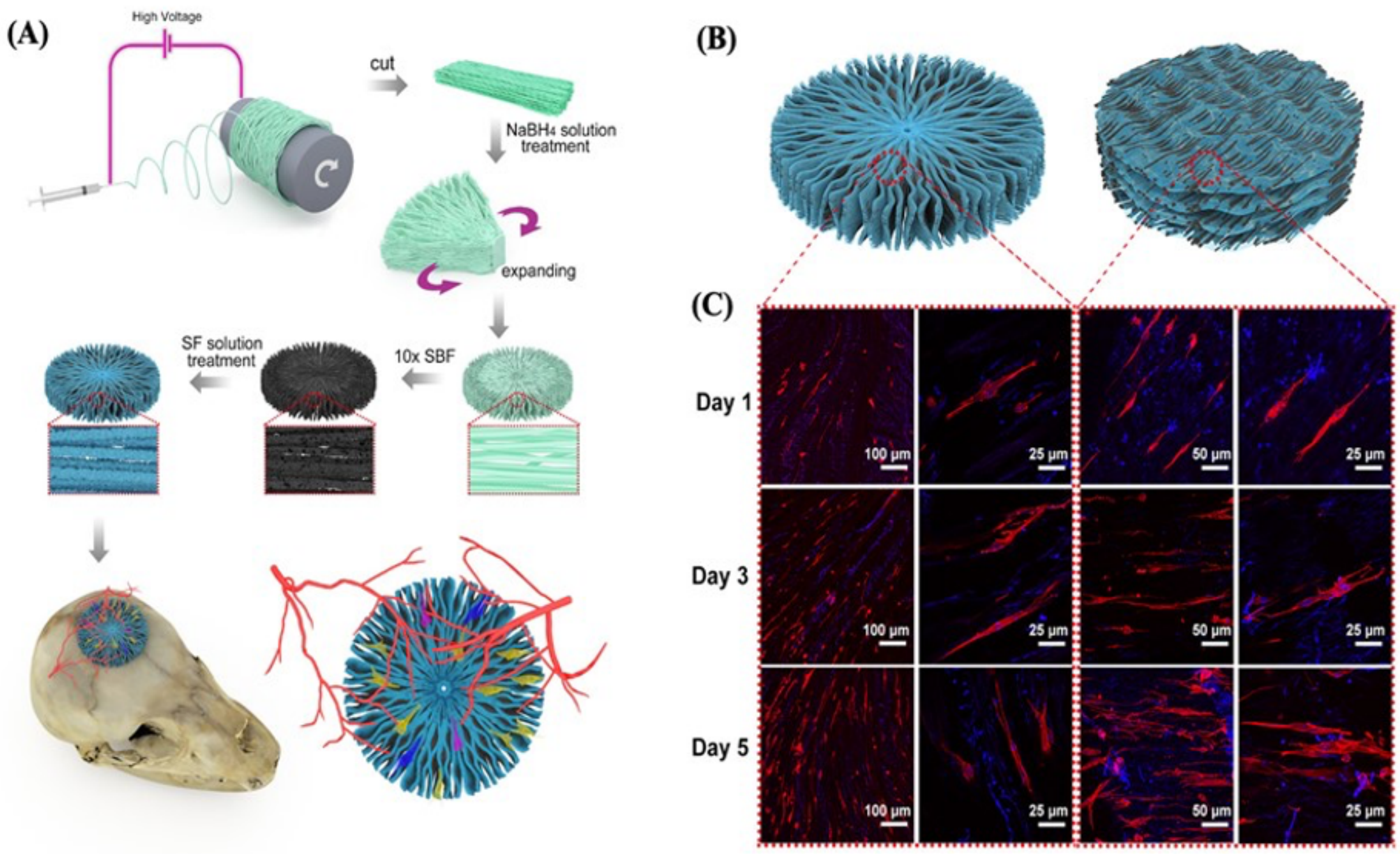

**Fig 2.** Schematic showing the process used to fabricate the 3D-radially aligned nanofiber scaffolds and it's the application for bone regeneration in vivo in a rat cranial defect model. (B) Schematic diagram showing the radially and laterally aligned scaffolds. (C) BMSCs proliferation and morphology on these scaffolds after seeding for 1, 3, and 5 days [31].

Incorporating equine bone-derived nanohydroxyapatite (EBNH) with PCL scaffolds has been found to boost osteogenic differentiation, especially in dental implants. Rahman et al. fused nanofiber with microspheres for controlled growth factor release, showcasing promising results for dental and bone tissue engineering [71, 72]. He et al. devised a method using metallic electrode plates to enhance the collection of nanofibers, leading to the creation of osteon-mimetic scaffolds. This method significantly influenced angiogenic differentiation and osteogenic protein production [67]. Scar tissue formation is a major concern in tendon healing. Faccendini et al.'s hybrid tubular scaffold demonstrates the potential to mimic the natural structure of tendons, promoting improved recovery after surgeries [73, 74]. Given its biocompatibility, cellulose is emerging as a potential scaffold material. Zhang et al.'s research on cellulose/cellulose nanocrystal nanocomposite nanofibers highlighted its potential in bone regeneration, especially when combined with BMP-2 [75]. Cristofaro et al. explored poly(butylene succinate) in scaffold design, emphasizing the scaffold's influence on cell differentiation [31]. Lastly, Ding et al.'s novel wet-collection electrospinning method showcases the benefits of fiber alignment and distribution for osteogenic differentiation [66].

### 2.2 Aligned nanofiber for neuronal tissue engineering

Neuronal tissue engineering is a multidisciplinary field that aims to develop artificial neural tissue in the laboratory for various research and therapeutic purposes. Though conventional treatment methods for nerve damage, such as grafting, are possible, they have major limitations due to the low availability of donors and low efficiency for repair. The primary objective of neuronal tissue engineering is to replicate the complex structure and function of the nervous system. This enables researchers to study neural development, model neurological diseases, and potentially restore or repair damaged neural tissue. One approach to advancing neuronal tissue engineering is the application of aligned nanofibers [33, 34, 76][45-47]. Electrospun-aligned nanofibers can have their mechanical properties, physical characteristics, and composition regulated through post-fabrication modification. For ideal neuronal tissue substrates, the direction of neurons and electrical conductivity are crucial factors. Aligned nanofibers aid in the growth of neurites and guide their spatial orientation in an organized manner [35-37]. Overall, aligned nanofibers are ideal for stimulating non-self-repairing neuronal tissue.

Biomimetic scaffolds created through electrospinning are being studied for their effectiveness in nerve tissue engineering. A research group led by Amini et al. used PCL, to collect electrospun aligned fibers with varying concentrations of lignin nanoparticles. The group found good cell viability and adherence of both rat pheochromocytoma (PC12) and human adipose-derived stem cells (hADSCs). The nanofiber scaffolds displayed a smooth surface, and the water contact angle decreased with higher lignin concentration. Moreover, the increase in lignin content highly stimulated axon growth of the neuron, while markers indicating cell differentiation were significantly upregulated. *In vivo* results demonstrated that scaffolds consisting of 15% lignin content had significantly enhanced neuron regeneration [33]. PCL can be coated with a wide range of biopolymers to improve the effectiveness of the scaffold. One strategy to promote neural repair is by providing stimuli for nerve regrowth. Zhang et al. developed a novel laser microdissection based axotomy model to study the effect of miR-coated PCL aligned fiber scaffold on axonal regeneration. The presence of miR, specifically miR132, miR222, and miR431, plays a significant role in protein synthesis. Delivery of miR to injured neurons has been shown to significantly enhance axonal regrowth. The axotomized dorsal root ganglia (DRG) showed significant axonal regrowth when cultured on the miR-coated fiber scaffold. Furthermore, the scaffold demonstrated longer neurite outgrowth post-recovery from sciatic nerve transection injury than untreated cells [34]. Nanofibers made solely of PCL have a few drawbacks, such as slow degradation rate and weak mechanical properties. To improve the characteristics of nanofiber scaffolds, researchers have studied the benefits of combining PCL with other polymers. Quan et al. developed a multi-step electrospinning process to fabricate nerve guide conduits using chitosan-incorporated PCL aligned nanofibers. Schwann cells, PC12 cells, and DRG were cultured on the fibers, monitoring Schwann cell markers (S100) and neurofilament marker (NF200) for each cell type. Additional mechanical testing of the polymer showed that the aligned nanofiber had better compressive performance as opposed to the randomly oriented fibers. *In vivo* experiments with right sciatic nerve defective mice, treated with aligned nerve conduit, revealed promotion for a longer axon had occurred in PC12 and DRG. An upregulation of activating transcription factor 3 (ATF3), cleaved caspase 3, and enhanced recovery of distal nerve were observed [76]. Stimulating the immune system through cells like macrophages can help trigger the recovery of neural tissue. Jia et al. conducted a study where they created aligned and random PLLA/PCL nanofibers to observe macrophage activity. The aligned scaffold encouraged macrophages to elongate along the fiber direction and produce a pro-healing macrophage phenotype (M2 type). In vitro results demonstrated that macrophages improved Schwann cell migration and proliferation. In vivo results on rat sciatic nerve defect models using differently oriented nerve guidance conduits revealed a higher presence of pro-healing macrophages in the group treated with aligned nanofiber. Additionally, the aligned group exhibited a 2-fold increase in Schwann cell infiltration and a 2.84-fold increase in axon numbers [38].

Maintaining the intrinsic functionality of neural networks is crucial for proper nerve cell interactions. Carbon nanotubes (CNTs) have gained attention for their surface nanotopography, which mimics the ECM. Electrical stimulation can significantly impact the potential of restoring nerve cell interactions. In studies conducted by Wang et al. and Hu et al., CNTs were incorporated to achieve this goal. Wang et al. developed a core-shell scaffold for peripheral nerve tissue engineering. The research group used a dry-wet electrospinning method to collect aligned conductive nanofiber yarns (NFYs) encapsulated with hydrogel. NFYs contained inside hydrogel were photocrosslinked to create a stable 3D microenvironment that mimics the native tissue structure (Fig 3 A). The NFYs consisted of PCL, silk fibroin, and CNTs, which induced proper neuron alignment of PC12 and DRG cells. NFY with all three diameters (25, 50, and 100 μm) showed excellent cell viability (Fig 3 B). The aligned scaffold supported cell growth and elongation, closely resembling the original epineurium layer [35]. Hu et al. used amine-functionalized multi-walled carbon nanotubes (MWCNTs) to enhance the electrical conductivity of PCL/GT nanofibers without affecting their biocompatibility. These MWCNTs were incorporated into the nanofibers to create random and aligned conducting scaffolds. The aligned conducting nanofibers significantly improved the differentiation of BMSCs to Schwann cells, enhancing peripheral axonal regeneration in both in vitro and in vivo experiments [77].

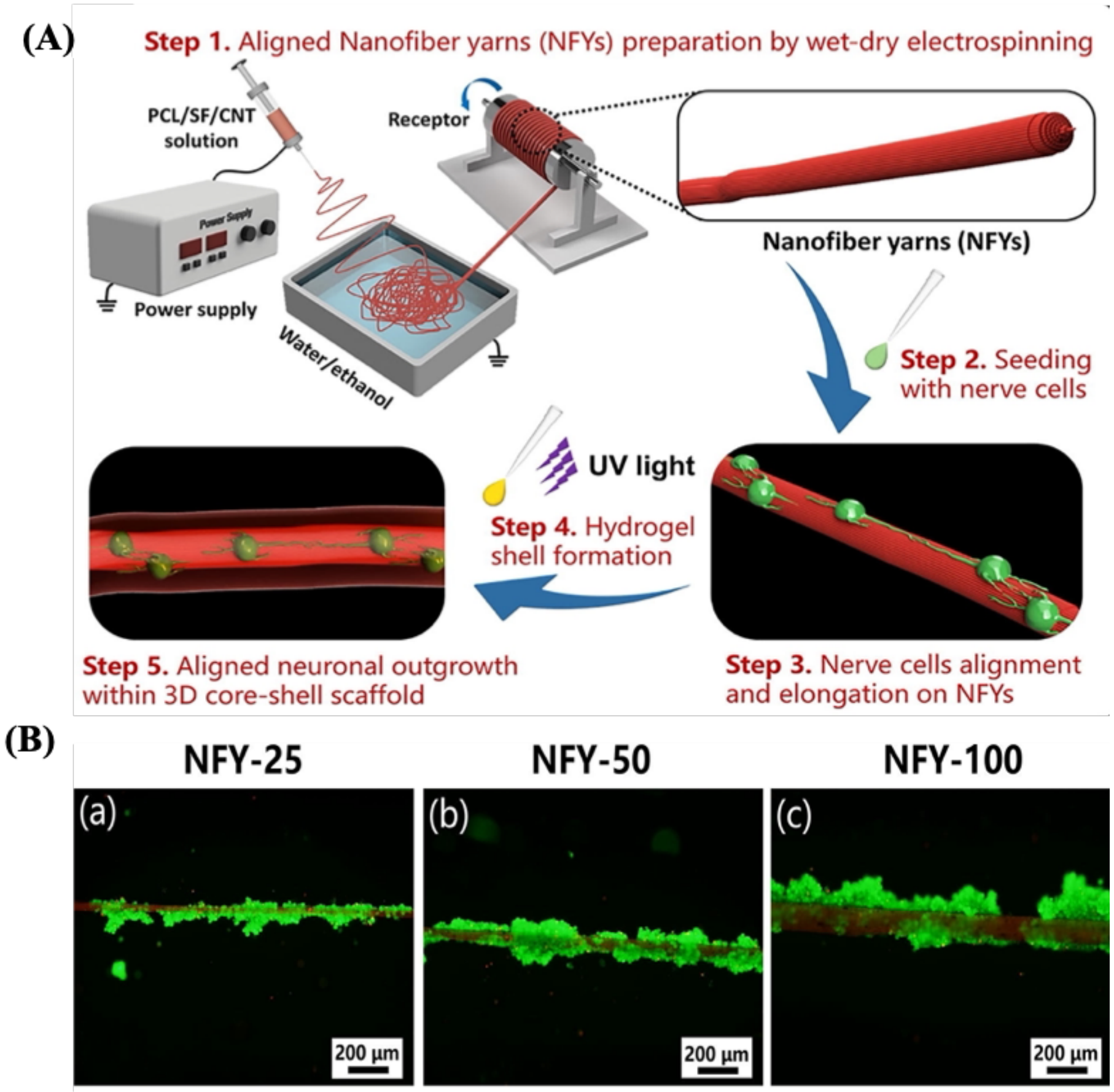

**Fig 3.** (A) Scheme for preparing conductive nanofiber yarn and promoting neuronal outgrowth within 3D core-shell scaffolds. (B) Live/dead stain of PC12 cell cultured on aligned NFY-25 (a), NFY-50 (b), NFY-100 (c) after culturing for 1 day [35].

Wang et al. used electrospun PLGA and MWCNTs to create a scaffold, which was then coated with poly-L-lysine. This composite scaffold had favorable electrical properties, thanks to the hydrophilic surface of the scaffold that enabled good cell attachment for PC12 cells. Additionally, DRG neurons were properly oriented along the fiber direction to promote neurite growth. When exposed to a 40mV electrical stimulus, PC12 cells and DRG neurons responded with long neurite length, high cell proliferation, differentiation, and myelination [39].

A noteworthy characteristic of PLGA is its potential for drug delivery. Tang et al. utilized this feature to create a specialized biofunctional scaffold made of LysoGM1-functionalized PLGA, which has the added benefit of promoting neuron growth and brain tissue regeneration. The aligned scaffold acts as a guide for neurite outgrowth and facilitates the organization of the neural network. In vitro testing showed that the scaffold stimulates synapse formation while maintaining neuronal viability. Moreover, in vivo studies conducted on rats with traumatic brain injuries revealed significant recovery when implanted with the aligned PLGA-LysoGM1 scaffold, as opposed to the blank control. The addition of LysoGM1 to the scaffold resulted in positive outcomes, indicating that other PLGA derivatives may also have favorable results [36]. In other studies, Cheong et al. incorporated PLGA and MAP-I, a bio-engineered fusion protein of mussel adhesive proteins (MAP) and IKVAV peptide, to induce nerve regeneration (Fig 4 A). They fabricated a scaffold using electrospun PLGA fused with MAP in an aligned orientation, which improved solubility and mechanical strength. In vitro studies using Schwann cells revealed better adhesion due to the integrin activity of ECM molecules, while a rat sciatic nerve defect model was treated with PLGA/MAP-I, resulting in speedy recovery and the formation of endoneurium structures and nerve fascicles. The author proposed faster proliferation, quicker differentiation and axonal growth could stimulate the regeneration process (Fig 4 B) [37].

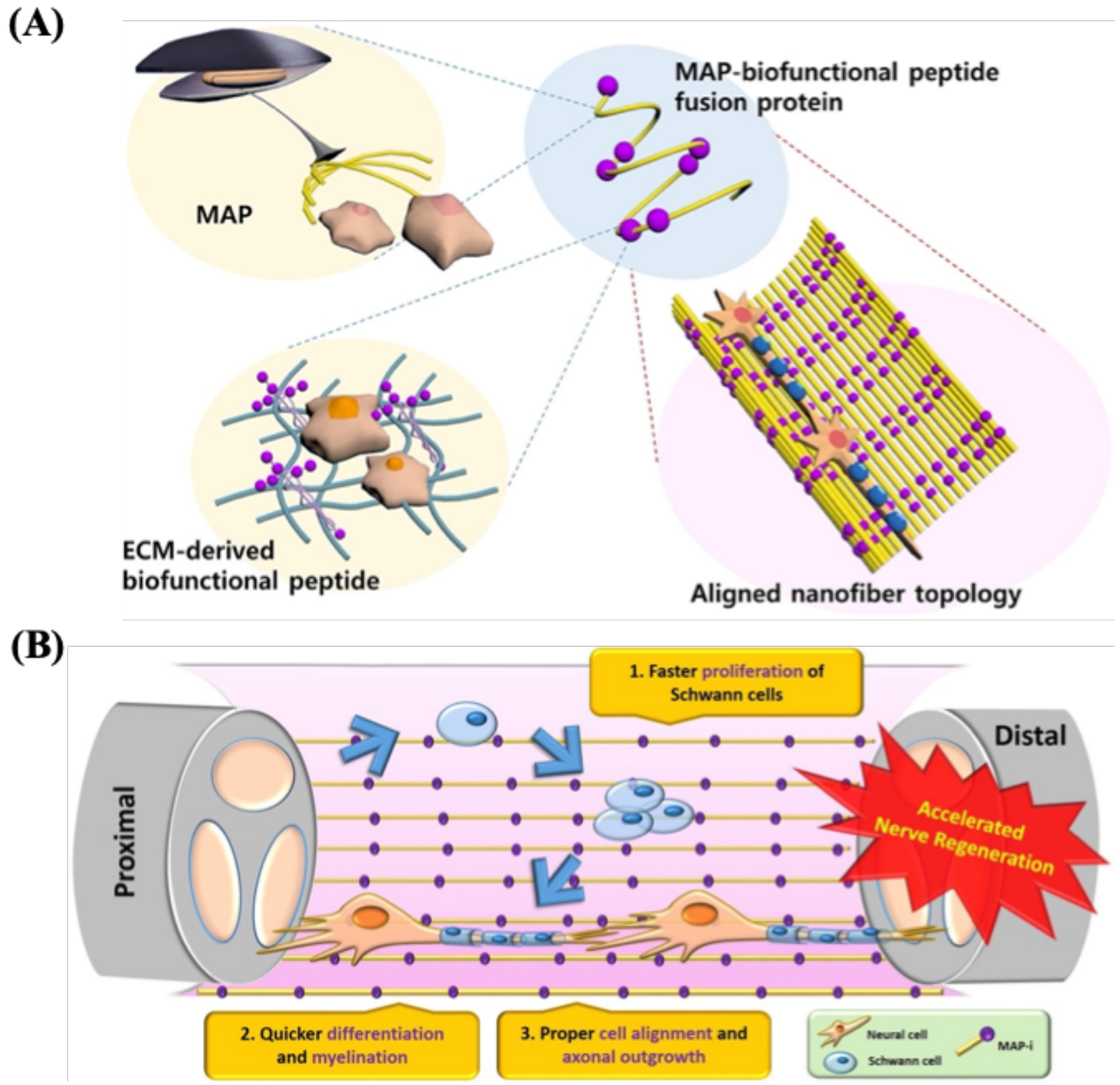

**Fig 4.** (A) Schematic representation of functional nerve regeneration on aligned nanofiber. (B) Proposed accelerated functional nerve regeneration process by MAP-i-incorporated aligned nanofiber conduit [37].

PLLAs are synthetic polymers particularly well-suited for neuronal tissue engineering because of their strength, biocompatibility, and biodegradability. A study by Barroca et al. found that electrical polarization could be used to create a platform for neural tissue regeneration. In vitro studies using neuroblastoma cell cultures indicated that retinoic acid triggered neuronal differentiation. Moreover, embryonic cortical neuron differentiation on poled-aligned nanofibers improved neurite outgrowth by 30% compared to its seeding onto non-poled-aligned nanofibers [40]. Furthermore, Ziemba et al. improved the functionality of PLLA fibers by modifying their surface. They did this by coating the PLLA fibers with silk fibroin (SF) through an immersion process in a coating solution. This method allowed for the self-assembly of SF to take place within the solution, inducing beta sheet assembly. In vitro tests showed that DRG cells were better able to attach to the SF-coated scaffold and exhibited longer neurite growth [41].

Aliphatic polyesters, such as PLLA, Poly (3-hydroxybutyrate-co-3-hydroxyvalerate) (PHBV), and polyhydroxyalkanoate (PHA), are known for their biodegradability. In a study by Hu et al., PHBV was electrospun in both a randomly oriented and aligned manner to culture, fibroblast growth factor (FGF2), and miR-218 treated ADSC. The FGF-miR-218 treated stem cells implanted into 10 mm transected rat sciatic nerves resulted in proper motor function restoration following the fiber alignment's direction, showing nerve damage recovery [78]. A study by Lizarraga-Valderrama et al. used a blend of poly(3-hydroxybutyrate) (P3HB) and poly(3-hydroxy octanoate) (P3HO) in a ratio of 25:75 to create PHA. They fabricated electrospun nanofibers used as guiding lumen structures for nerve conduits. NG108-15 neuronal and RN22 Schwann cells showed that PHA blends nanofibers improved growth and assisted with neuron-aligned distribution. Additionally, a direct relationship was observed between fiber diameter and neuron growth. Co-culture of the two cell lines drastically increased neuronal growth [42].

Aromatic polymers are excellent organic materials with unique mechanical and electrical properties, and are chemically stable. Among the polymers that have potential applications in neuronal regeneration, polyethersulfone (PES), polyurethane (PU), and polypyrrole (PPy) are noteworthy. Ghollasi et al. created a nerve regeneration scaffold by using aligned laminin-immobilized PES, and cultured human-induced pluripotent stem cells (hiPSC). They produced two different types of fiber alignment, random and aligned, by means of electrospinning, followed by $O_2$ plasma treatment for laminin binding. The functionalized scaffolds showed improved neuronal gene expression compared to the control group. Furthermore, in the aligned functionalized fibers, hiPSCs revealed neurite growth towards the direction of the fibers. The aligned pure fiber exhibited increased expression of neuron-specific enolase, Tuj-1 (axonal marker), and microtubule-associated protein 2 (dendritic marker). Overall, the functionalized aligned PES fiber's hydrophilicity and biocompatibility demonstrated good adhesion, proliferation, and differentiation of hiPSCs into neurons [79]. Shrestha and colleagues developed a hybrid material made of polyurethane (PU), silk fibroin, and functionalized multiwalled carbon nanotubes (fMWCNTs) via electrospinning. The material was found to be hydrophilic and biocompatible, with significantly improved electrical conductivity due to the addition of fMWCNTs. In vitro experiments using Schwann cells and PC12 showed high proliferation and neurite outgrowth, respectively. The cells along the aligned fibers grew in an axis-aligned manner, which was not observed in the randomly oriented scaffold [80]. Elashnikov et al. also investigated another conducting polymer, PPy. In their study, they found that electrical stimulation from PPy induced nerve regeneration. The research group created nanofibers of PPy-coated cellulose acetate butyrate (PPy/CAB) with both uniaxial and random orientations. The scaffold's topography improved the wettability of the nanofibers. In vitro studies cultured neuroblastoma cells onto the PPy/CAB nanofibers, which were stimulated by 100mVmm-1 at 1Hz pulses revealed that the uniaxially aligned and stimulated nanofibers had enhanced cell adhesion and neurite outgrowth [43].

Natural polymers like chitosan and silk fibroin are excellent biomaterials that promote neuronal cell adhesion and proliferation. Multiple stimuli can be employed to initiate neural tissue engineering, which can be incorporated into a single scaffold for long-term release. Li et al. utilized silk fibroin fiber and functionalized with laminin, using carbodiimide coupling, a type of covalent binding. In vitro, evaluations of cultured PC12 cells showed that this scaffold stimulated directional axonal extension and provided other signals for neurite outgrowth. The observed neurites were similar in length to native neurons, indicating a good use of silk fiber as a nerve graft [81]. In a study of Nune et al. silk fibroin mixed with melanin to produce randomly oriented and aligned conducting composite scaffolds. Cell culture of SH-SY5Y cells showed that they adhered properly and had good viability. The composite was highly antioxidant, removing free radicals. Additionally, the aligned polymer prompted the differentiation of neuroblastoma cells into neurons, guiding their proper orientation [44]. In addition, Rao et al. have developed a hydrogel made of aligned chitosan fiber, which was created through liquid electrospinning and then manipulated using mechanical stretching processes. The hydrogel was then grafted with RGI and KLT peptide. RGI is a type of brain-derived neurotrophic factor, while KLT mimics the function of VEGF. These bioactive peptides are essential in aiding nerve regeneration and angiogenesis, respectively. In vitro tests using Schwann cells on a scaffold showed cell reorientation and proliferation, resulting in increased production of neurotrophic factors. The hydrogel generated nerve regeneration and induced vascular penetration when injected into rat sciatic nerve defects. After 12 weeks, the rat model showed functional recovery at the injury site [82].

### 2.3. Aligned nanofiber for cardiac tissue engineering

Cardiac tissue engineering is a specialized field of regenerative medicine that focuses on developing functional and contractile cardiac tissue. Its primary application lies in cardiac research and regenerative therapies. The main objective of cardiac tissue engineering is to create effective strategies to repair or replace damaged or diseased heart tissue caused by myocardial infarction (MI) or congenital heart defects. Cardiomyocytes (CMs), the specialized cells of the heart, are incapable of division for repair in heart disease. Therefore, the application of biomimicking substrates, such as aligned nanofiber scaffolds, holds great potential to help restore functional cells in the cardiac tissue. The aligned fibrillar structure provides topographic signals for cell orientation in the cardiac tissue, which has inherent anisotropic textures and functions [83, 84]. These scaffolds provide mechanical stability, biocompatibility, and microenvironment for cellular communication [85, 86]. The constant movement of cardiac tissue and complex cellular networks are crucial for proper function.

Aligned nanofibrous scaffolds have gained attention in cardiac tissue engineering [45, 87]. Researchers have found that human-induced pluripotent stem cell-derived cardiomyocytes (hiPSC-CMs) bear a strong resemblance to immature CMs with regard to their morphology and function. A team led by Li et al. has developed high-quality cardiac tissue-like constructs (CTLCs) by seeding hiPSC-CMs onto aligned PLGA polymer (Fig 5 A). The nanofiber was fixed into a silicon frame for convenience, and the cultured cells exhibited a highly organized, complex, and ideal elongation in vitro. The cells significantly increased cardiac tissue-specific markers such as β-MHC on aligned fibers compared to flat surfaces. α-Actinin-positive sarcomeres and cardiotropin (cTnT)-positive myofilaments were also well-defined and positioned along the aligned fiber compared to random and flat surfaces (Fig 5 B). In an MI rat mouse model, CTLCs on damaged heart tissue showed positive results for repairing the tissue (Fig 5 C). The hiPSC-CMs expressed human TnT in the heart tissue 14 days after transplantation [14].

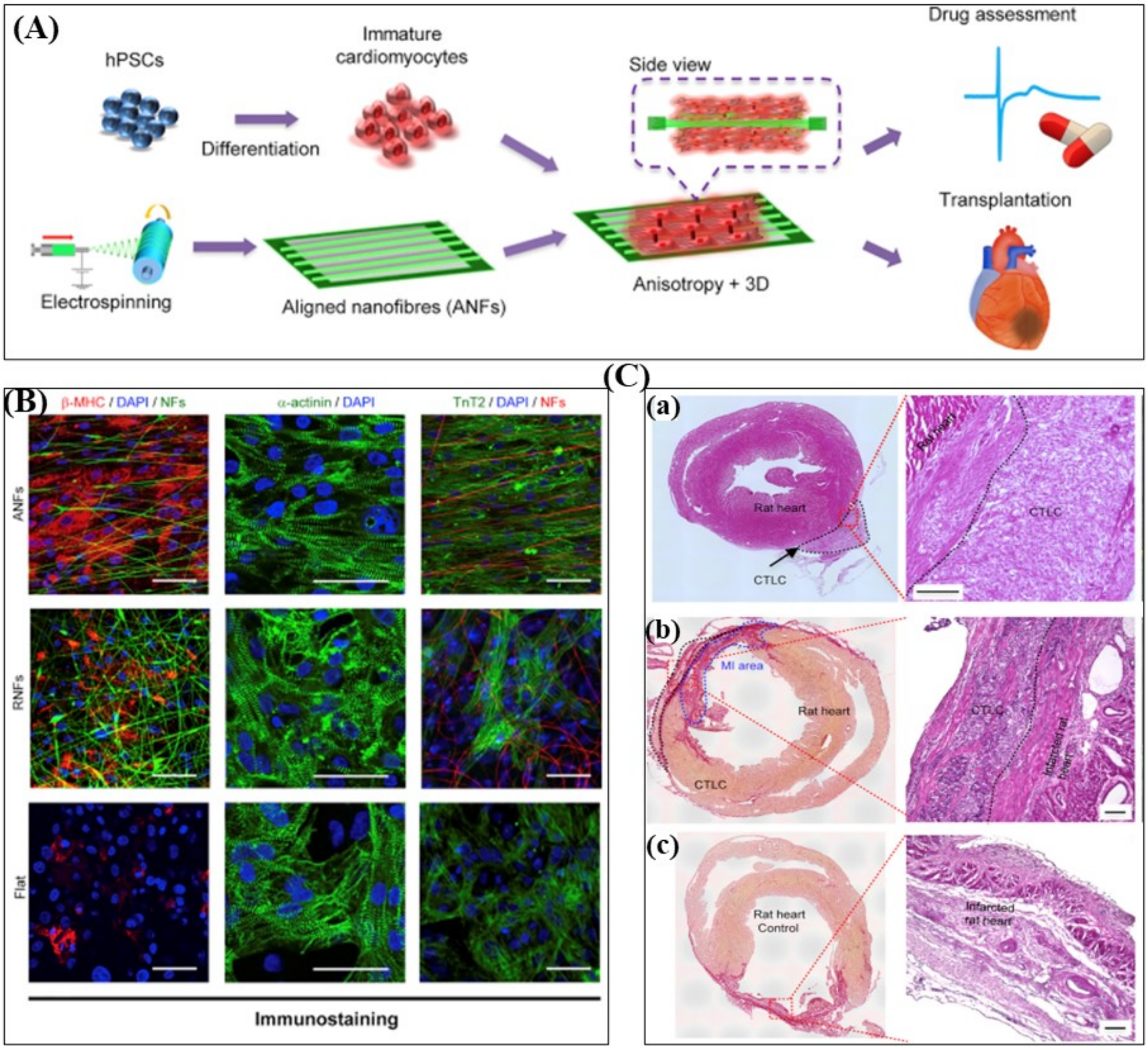

**Fig 5.** (A) Schematic illustrates the detail experimental approach. (B) Immunostaining images of b-MHC (red), and a-actinin and cTnT (green). Cardiomyocytes were cultured on different substrates: aligned nanofibers (ANFs), random nanofibers (RNFs), and gelatin-coated flat substrate (Flat); nanofibers were fabricated with green or red dyes added before cell seeding. Scale bar, 50 mm. (C) Histological sections of CTLCs after being transplanted on the rat heart without (a) or with myocardial infarction (b) and MI heart with acellular nanofiber (c) [14].

Understanding the contraction force of cardiomyocytes or heart tissues is crucial in comprehending their function. Electrospun scaffold can be used to create beating myocardial constructs. In a study by Eom et al., hiPSC-CMs exhibited functional multiaxial contraction on an aligned-random heterogeneous PCL nanofiber mat. The mat had distinct areas with either aligned or random orientation, allowing CMs to be guided in the correct orientation through directional cues by the aligned section. The random orientation of the mat supported stability. The thin nanofiber mat exhibited a contraction velocity about 20 times faster than the thick nanofiber mat and flat surface [46]. Similarly, in a study by Ding et al., iPSC-derived cardiac progenitor cells (CPCs) exhibited contraction results when placed on the aligned PCL nanofiber. The surface of the scaffold was designed in a way that allowed for cell adhesion and triggered cell differentiation into functional CMs by inhibiting Wnt and BMP signaling. After 14 days of treatment with an inhibitor, the cells displayed spontaneous contraction. The functionality of the CMs was confirmed through additional evaluations of cTnT expression and intracellular $Ca^{2+}$ oscillation. The aligned scaffold provided an ideal microenvironment for the proper maturation of iPSC-CPCs, resulting in efficient contraction [47]. In addition, researchers found that treating PCL fiber with oxygen plasma and sodium dodecyl sulfate solution improved the ability of hiPSC-CM cells to stick to the surface. The cells also displayed a more elongated shape aligned with the direction of the fibers, compared to cells grown on random fibers or a culture plate. Additionally, the scaffold enhanced sarcomere organization, spontaneous synchronous contraction, and the expression of mature genes in the CMs [48].

Achieving an anisotropic cardiac structure can be challenging. Wu et al. developed a conductive nanofiber yarns network (NFYs-NET) using a weaving technique. The scaffold comprises PCL, silk fibroin, and CNT encapsulated inside a GelMA hydrogel shell after photo cross-linking. The structure closely resembles the intricate interwoven nature of native cardiac tissue (Fig 6 A). The scaffold enables CMs to elongate and align across each layer by regulating cellular orientation. Multiple cardiac markers were highly expressed, such as sarcomeric alpha-actinin and gap junction protein connexin-43 (CX43) in 50μm diameter NFY (NFY-50) (Fig 2 B). This scaffold is promising for 3D cardiac anisotropy therapeutics [45].

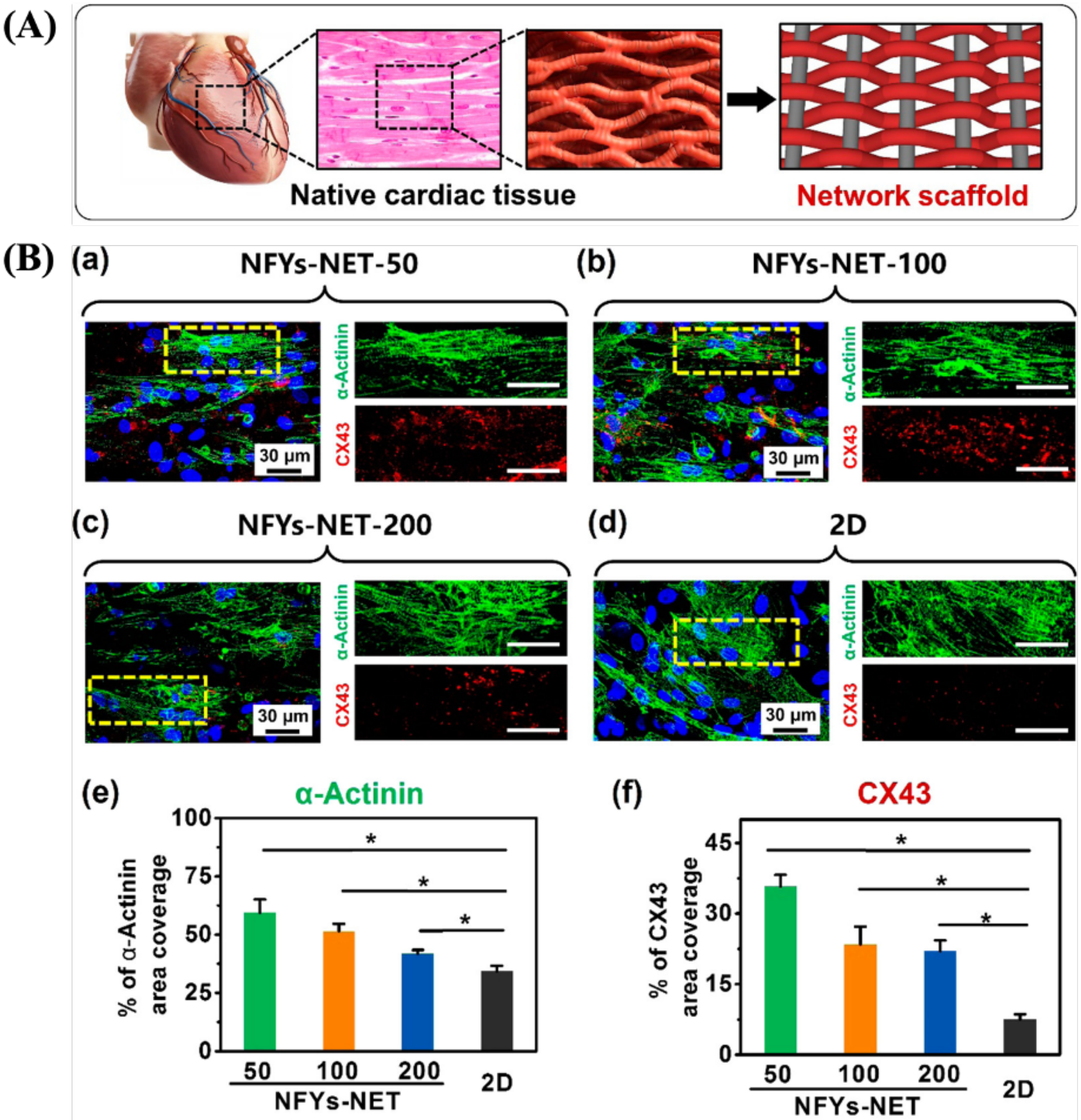

**Fig 6.** (A) Designing an NFYs-NET scaffold, having a network structure would be beneficial for cardiac tissue regeneration (B) Immunofluorescence staining of sarcomeric α-actinin (green), nuclei (blue), and CX43 (red) revealed that there were phenotypic differences between cardiac tissues on NFYs-NET scaffolds (a−c) and the 2D control group (cultured for 8 days) (d). α-actinin immunostaining demonstrated a well-organized sarcomeric α-actinin and sarcomere alignment on NFYs-NET scaffolds, while CX43 immunostaining revealed a better interconnected sarcomeric structure with robust intercellular junctions on NFYs-NET scaffolds compared to the 2D group. The scale bar was 30 μm. The relative percentage of area coverage by α-actinin (e) and CX43 (f) was quantified on NFYs-NET scaffolds and on the 2D group, respectively [45].

One of the therapies for repairing damaged heart tissue is to create patches that closely resemble the properties of the heart. These patches should be able to stretch, contract, and maintain their natural elasticity. Cesur et al. conducted a study where they produced patches made of random PLA, PLA/PEG (Polyethylene glycol), PLA/PEG/Collagen, and aligned PLA/PEG/Collagen. The aligned nanofiber patch had the highest tensile strength (5.90 ± 4.00 MPa) and degradation ratio compared to the random fiber. Furthermore, the human cardiomyocyte cell line (H9C2) grew uniformly and aligned with the aligned PLA/PEG/Collagen fiber [49]. A cardiac patch is used as a scaffold to transfer stem cells or differentiated cardiac cells in order to build viable and functional heart tissue. A promising patch made of PLA/chitosan nanofiber was found to be seeded with primary CM for regenerating myocardia. An aligned scaffold made of PLA/chitosan nanofiber with an ideal ratio of 7:1 was found to have enhanced hydrophilicity and a small fiber diameter. The aligned fiber had better mechanical properties and promoted biocompatibility while stimulating cell elongation of primary CMs. Additionally, it was found to stimulate the expression of sarcomeric alpha-actinin and contractile protein (troponin I) [50].

When engineering cardiac tissue, selecting appropriate biomaterials is a significant challenge. However, in recent years, polyurethane-based (PU-based) scaffolds have shown promise as a solution [84, [88]. A composite scaffold of PU, chitosan, and CNTs has proven effective in regenerating infarcted myocardium. The scaffold includes randomly oriented and aligned scaffolds replicating the ECM structure. PU provides structural stability, chitosan ensures biocompatibility, while CNTs add electrical conductivity properties. H9C2 cells were cultured onto the nanofibers, demonstrating excellent cell proliferation and viability on the aligned composite scaffold [85]. PU composites are widely used due to their ability to customize mechanical properties, good degradation rates, and general biocompatibility. Anisotropic PU and ethyl cellulose scaffolds with aligned nanofibers have demonstrated high mechanical strength, making them ideal for supporting contractile cardiac tissues. The in vitro experimental results have shown that these scaffolds promote the proliferation and viability of cardiac myoblast H9C2 cells [51]. Chen et al. developed an aligned, randomly orientated degradable polar/hydrophobic ionic polyurethane (D-PHI) and polycarbonate polyurethane (PCNU) composite scaffold. D-PHI, a material selected for its immunosuppression activity, was used to deliver cardiomyocytes into the tissue as a potential treatment for myocardium repair. After one week, introducing hiPSC-CMs with the scaffold resulted in a similar level of cardiac troponin-T but a much higher level of ventricular myosin light chain-2 compared to the cells cultured on a tissue culture plate. Additionally, the cells on the aligned scaffold were beating synchronously, similar to the ones cultured in the control group [86]. Similarly, PEUU blended with gelatin scaffolds-maintained shape during cyclic deformation, while CMs cultured on the aligned scaffold proliferated well due to its anisotropic characteristic [52].

## 2.4. Aligned nanofiber for muscle tissue engineering

Muscle tissue engineering is an exciting and promising field that seeks to restore muscle tissue function for various purposes, such as regenerative medicine, drug testing, and disease modeling. One of the ways to enhance muscle tissue engineering is by utilizing aligned nanofibers as a scaffold [89, 90].

Nanofibers that are aligned provide a biomimetic environment for muscle cells, also known as myocytes, to mature, orient, and function in a way that is similar to the native muscle tissue. The alignment of the nanofibers mimics the complex organization of muscle fibers, which in turn promotes the formation of intricate muscle tissue constructs [91]. The aligned nanofiber scaffolds can be made from various biocompatible materials such as synthetic polymers (e.g., polycaprolactone, poly(lactic-co-glycolic acid)) or natural polymers (e.g., collagen, fibrin) [53, 54, 92]. Subsequently, the muscle cells (myoblasts) such as primary muscle cells or stem cells can be induced to differentiate into muscle cells on these fibers [55, 93, 94]. To induce muscle tissue formation, differentiation can be promoted by providing specific growth factors and culturing the cells in conditions that mimic the biochemical and mechanical cues of native muscle tissue.

Poly(ε-caprolactone) (PCL) is widely used in muscle tissue engineering. It has already been approved by the FDA. However, it has some limitations due to its hydrophobic nature and low degradation rate, which can negatively affect cell interaction and post-implant resorption. To overcome these limitations, PCL has been combined with other biopolymers in scaffold fabrication, resulting in improved properties [95-97]. For example, Cai et al. blended PCL with collagen to enhance biocompatibility. The study demonstrated that Schwann cells promote myoblasts' myogenic differentiation when cocultured on aligned PCL-collagen I nanofibers. The myoblast upregulated the myogenic markers such as myosin heavy chains (MYH2) and myogenin (MYOG) after 28 days of culture with the presence of Schwann cells and adipogenic mesenchymal stromal cells (ADSCs). The study highlights the interactions between Schwann cells and other cell types, potentially opening new avenues for tissue engineering and regenerative medicine applications [53]. A recent study has found that when primary myoblasts and mesenchymal stromal cells (MSCs) were co-cultured on a combination of PCL and collagen I in a serum-free environment, it resulted in the highest upregulation of myogenic markers. The nano scaffolds provided a conducive environment for the growth and proliferation of both cell types under serum-free conditions. The upregulation of myogenin markers was 5.2-fold in ADSC/myoblast and 2.1-fold in BMSC/myoblast. Furthermore, the formation of multinucleated myotubes significantly enhanced myogenic differentiation. The use of serum-free conditions eliminates the risk associated with animal-derived serum, making the nano scaffolds more suitable for therapeutic purposes in potential clinical translation [56]. Witt et al. found a similar result when studying the co-culture of MSCs and myoblasts. The study involved using aligned electrospun PCL collagen-I nanofibers and the addition of hepatocyte growth factor (HGF) and insulin-like growth factor-1 (IGF-1) to differentiate myoblast cells into mature muscle cells. The combination of HGF and IGF-1 stimulation positively impacted the differentiation of both MSCs and myoblasts towards a myogenic lineage. By facilitating the expression of muscle-specific markers (desmin, myocyte enhancer factor 2, MYH2, and alpha-sarcomeric actinin), the growth factors aided in forming functional muscle fibers in the scaffold. The study concludes that by providing the appropriate biochemical cues, HGF and IGF-1 can enhance the differentiation of MSCs and myoblasts into mature muscle cells, which is crucial for successfully generating functional muscle tissue in vitro [55].

The combination of PCL with decellularized muscle ECM led to improved tensile mechanical properties of the electrospun scaffold. Using an aligned nanofiber scaffold resulted in a significantly higher myosin heavy chain (MHC): collagen ratio compared to untreated volumetric muscle loss (VML) injured models 28 days post-implantation. The group treated with the blend showed significantly higher upregulation of myogenic markers such as myogenic differentiation factor-1 (MyoD) and myogenin than the control group. The aligned structure of the nanofibers facilitated cell organization and supported the formation of new muscle tissue in vivo [54]. Another nanofiber blend of (poly(l-lactide-co-caprolactone) P(LLA-CL) and collagen, prepared as nanoyarn with a topological structure to native collagenous fiber, showed promising results for stress urinary incontinence sling when cultured with placental stem cells (PSCs). The cells adhered well to the nanoyarns, exhibited good viability and proliferation, and the myoblast-nanoyarn construct resembled muscle tissue. It increased the expression of myoblast markers such as α-SMA, desmin, and collagen type 1, 3 on the scaffold [92]. Likewise, a blended nanofiber yarn made of PCL, silk fibroin, and polyaniline was developed using a dry-wet electrospinning method. This yarn showed excellent biocompatibility, alignment, and elongation. Myoblasts enhanced MHC expression when cultured on nanofiber yarn/hydrogel core-shell scaffolds. Furthermore, the scaffolds maintained the mechanical properties needed for functional muscle tissue. The hydrogel shell creates a bioactive environment for nutritional exchange, facilitating myoblasts' differentiation into mature muscle cells [58].

Modifying nanofiber surfaces is an effective way to enhance cell behaviors. In tissue engineering, the surface engineering of scaffold materials plays a crucial role since these properties determine cellular behaviors such as adhesion, proliferation, migration, and differentiation [98, 99]. In a study by Ku et al., the surface of PCL nanofibers was modified with PDA, which resulted in improved behavior of skeletal myoblasts. The surface modifications with PDA exhibited a higher fusion index (27%) and maturation index (38.4%) than those without modifications, which recorded 17.1% and 14.1%, respectively. The PDA-modified PCL nanofibers significantly increased MHC expression and myoblast fusion [94]. The PCL fiber undergoes surface modifications through PDA treatment, which is then combined with PEDOT: PSS to create an electroconductive material. These matrices aim to mimic the structural and functional features of native muscle tissue, including its capacity to transmit electrical signals that are essential for muscle contraction. This supports a conducive environment for muscle cell adhesion, proliferation, and differentiation. Additionally, the matrices' electroconductive properties facilitate the transmission of electrical signals, which further promotes muscle cell alignment and maturation [59]. Combining surface modifications with surface topography can significantly increase myoblast differentiation. In a study by Zhang et al., they coated PCL nanofibers with gold nanoparticles to provide electrical stimulation, enhance alignment, and promote the maturation of myotubes. The spacing between microgrooves also significantly impacted the elongation and orientation of myotubes. When the microgrooves were spaced at 200μm, fabricated by a gold nanolayer, H9c2 myoblast cells showed highly aligned behavior compared to the 300μm size (**Fig 7 A B**). Additionally, rolling the scaffolds allowed the nanofibers to resemble aligned microscale basal lamina (**Fig 7 C)** [60]. The topographical cues provided by the scaffolds influenced cell alignment, elongation, and organization, which are important characteristics of mature muscle cells. In Cha et al.'s study, it was found that micropatterned PEG hydrogel on PCL nanofiber behaved differently for myogenic differentiation. The mature myogenic marker MHC was expressed at a low level in the random pattern, while it was highly expressed in the aligned nanofiber [100].

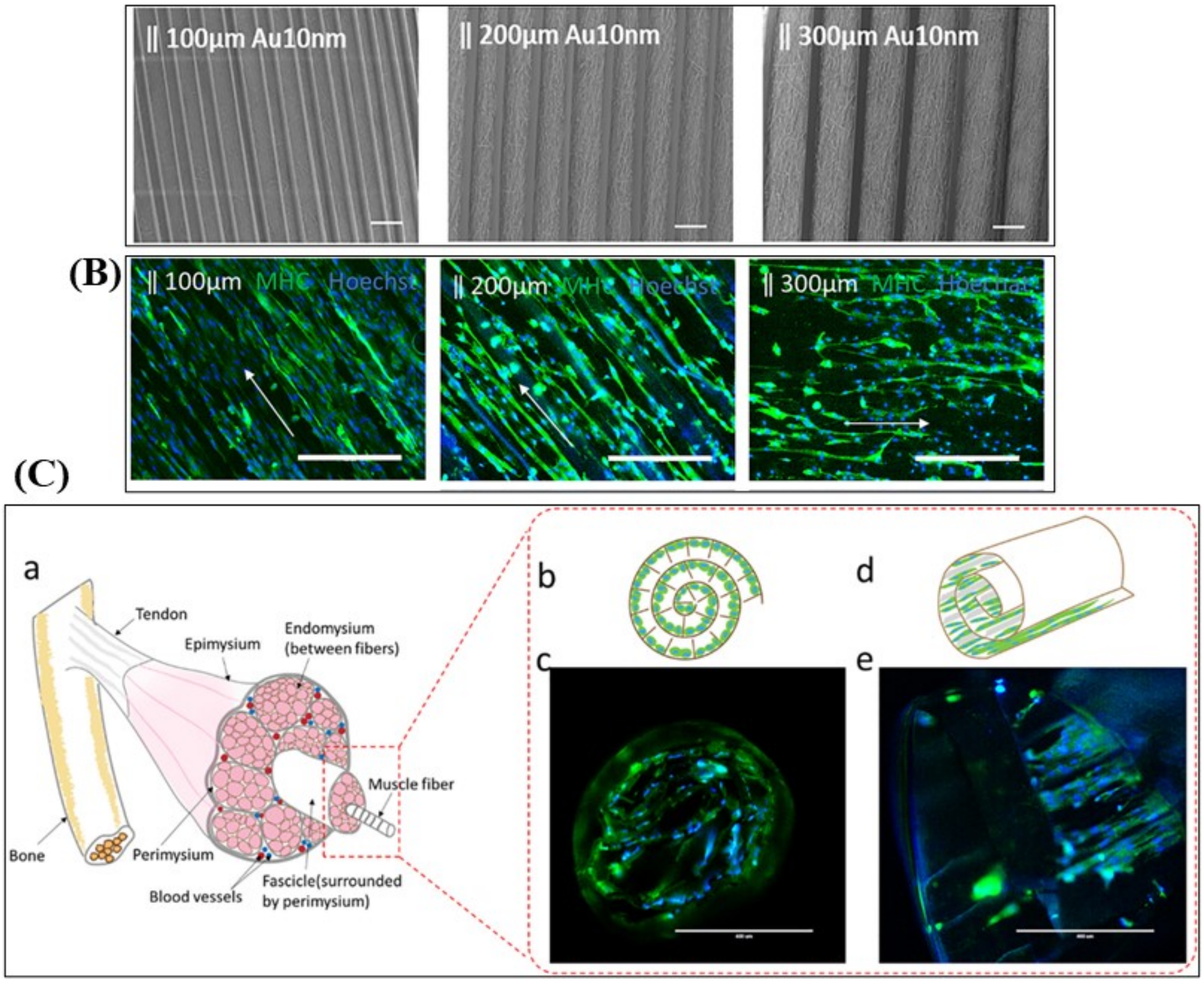

**Fig 7.** (A) The images captured using a scanning electron microscope depict a scaffold with microgrooves spaced at 100 μm, 200 μm, and 300 μm apart. Scale bar = 200 μm. (B) Cell alignment and myotube formation after cultured for 7 days, the scaffolds with spacing of ǁ100 μm, ǁ200 μm, and ǁ300 μm. (C) The skeletal muscle tissue was schematically represented, and the rolled 3D scaffold was observed from the lateral and cross-sectional view after being cultured with H9c2 cells for 7 days. Myosin heavy chain MHC (green) Hoechst (blue). Scale bars = 400 μm [60].

Cell electrospinning (CE) is a technique that comes with many benefits. It provides micropattern guidance, even cell distribution, and enhanced nutrient accessibility in 3D. By utilizing natural polymers and living cells, CE makes it possible for cells to grow via micro/nanofibrous structures. In a study conducted by Yeo et al., they combined a micro/nano-hierarchical scaffold with cell-printing of myoblasts. This resulted in favorable myoblast differentiation and improved cell alignment. The bioink used, which was composed of polyethylene oxide (PEO) and alginate on PCL struts, was laden with myoblasts. It showed rapid cell proliferation and differentiation when printed with an aligned pattern. Compared to random and no pattern scaffold, the aligned scaffold exhibited increased myotube number and length. After 20 days of culture in aligned pattern, myogenic genes (MyoD, myogenin, and troponin T) were significantly enhanced [61]. In their later study, researchers used C2C12 myoblast-laden alginate/PEO bioink to electrospin an aligned nanofibrous bundle onto a PCL strut in an anisotropic manner. The myoblast cells demonstrated a highly arranged and multinucleated morphology inside the fiber. Additionally, the myogenic markers, such as MHC, myotube formation and fusion, were higher in the cell-electrospun fiber compared to those without an aligned micropattern [101]. Furthermore, the researchers found that the micropatterned PCL strut increased mechanical stability and induced myotube formation. The combination of alginate and PEO nanofibers helped align myoblasts. The scaffold's micro/nanoscale hierarchical topography led to myogenic differentiation, with the highest MHC marker expression observed in alginate-electrospun PVA-leached PCL (patterned PCL strut) compared to alginate-coated PCL (non-patterned PCL strut) and alginate-electrospun PCL (non-patterned PCL strut). The patterned strut resulted in the longest myotubes (225.3 ± 51.4 μm) due to the synergistic topographical cue and efficient cell-to-cell interaction. Additionally, the patterned strut upregulated myogenic genes (MyoD, Myog, MHC, and TnT) after 28 days of culture. The collagen strut also showed similar MHC maturation index and length to the PCL strut. This study highlights the potential of alginate/PCL scaffolds with nano/microscale topographical design to induce myoblast alignment and myogenic differentiation [62]. Nanofibers loaded with single cells have several advantages, such as 3D accessibility to nutrients and cell-to-cell/matrix interactions. However, it takes a long time for the cells to interact with each other. To address this issue, Yeo et al. used spheroids-laden nanofibrous structures to achieve a high degree of alignment and differentiation of myoblasts. The spheroids-laden alginate/PEO nanofiber on the PCL strut effectively promotes myoblast alignment, which is achieved by the structural cues provided by the nanofibers guiding the organization of the cells. The aligned scaffold exhibited higher maturation and myotube formation [102].

Biomolecules can be used in scaffolds to achieve sustained delivery, and their use can be optimized for this purpose. In a study conducted by Moghadasi et al., the myogenic differentiation of hMSCs was investigated. The study evaluated the impact of surface topography and sustained release of TGF-β1. The results showed that hMSCs expressed smooth muscle cell markers at a higher intensity in PCL/PLLA nanofibers coated with TGF-β1 loaded chitosan nanoparticles 15 days post-culture, compared to the culture plate. The sustained release of TGF-β1 enhanced the differentiation of hMSCs compared to bolus delivery. The authors also explained the mechanisms behind TGF-β1 release from PCL/NP(TGF)-PLLA nanofibrous scaffold, which involves chitosan degradation and diffusion through the PCL nanopores [63]. Researchers have developed aligned nanofibers composed of RGD peptide-displaying M13 bacteriophage,

PLGA, and graphene oxide (GO) to promote myogenesis, which is the formation of muscle tissue. The RGD peptide, known for promoting cell adhesion, was displayed on the surface of the bacteriophage. The inclusion of GO in the nanofibers provided electrical conductivity, which has been shown to enhance myogenic differentiation. The aligned structure of the nanofibers mimicked the natural alignment of muscle fibers, further promoting myogenesis. The nanofiber sheets also upregulated myogenin and MHC gene expression [92]. Cheesbrough et al. took advantage of the hyperelastic properties of the P(EDS)UU-POSS nanofibers and prepared nanofiber sheets (Fig 8 A). These nanosheets had unique optical properties, which enabled control over the behavior of human induced pluripotent stem cell (iPSC)-derived skeletal myofibers through the projection of blue light and inducible Pax7 (iPax7) (Fig 8 B). The nanofibers facilitated the organization of myofibers in a specific direction and enhanced their contractile properties by over 200% and specific force by over 280%. It also promoted an increase of approximately 32% in myotube fusion and approximately 50% in myofiber maturation (Fig 8 C) [64]. Silk fibroin is a natural elastic polymer that is hydrophilic and has found extensive use in various biomedical tissue engineering applications. In a study conducted by Manchineella et al., they created a conductive fiber by combining silk fibroin with melanin. The resulting fiber not only supported the growth of mouse myoblast C2C12 cells but also led to better differentiation into aligned high aspect ratio myotubes, as compared to random films. Additionally, the scaffold exhibited impressive antioxidant properties and effectively reduced intracellular ROS levels [93].

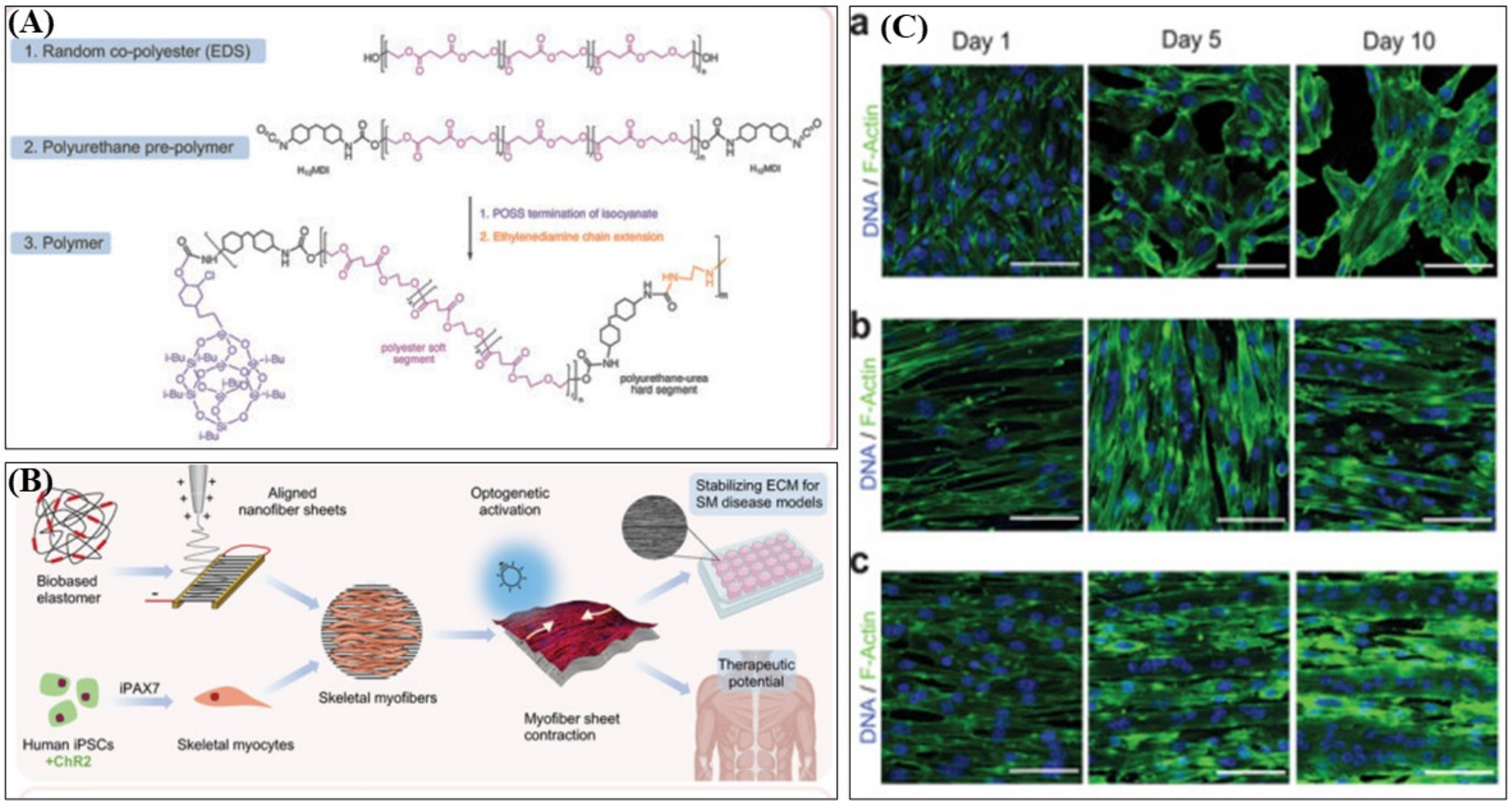

**Fig 8.** (A) The process of biobased elastomer synthesis involves several steps, including the molecular structure of polyester polyol EDS, prepolymer polyurethane, and P(EDS)UU-POSS copolymer nanohybrid. (B) The schematics of biobased elastomer nanofibers are used to guide iPSC-derived differentiation of skeletal myofibers. Aligned nanofibers are created by applying an electrostatic field between two aluminum electrodes. The differentiation of ontogenetically controlled skeletal myocyte progenitors from iPSCs is achieved using iPax7 forward programming (C) The morphology of Myotube and differentiation on elastomer nanofibers are also studied in this research. C2C12 myoblasts stained for F-actin (green) and DNA (blue) were grown on glass coverslips (control) and on coverslips coated with randomly oriented and uniaxially aligned P(EDS)UU-POSS electrospun nanofibers, on days 1, 5, and 10. The scale bars used are 100 µm [64]. EDS, ethylene-diethylene succinate; iPSC, human induced pluripotent stem cell.

## 3. Conclusion and perspectives

Electrospinning is a widely used technique for producing nanofibers in the range of nano and micrometers. When the electrospinning process is controlled to align the nanofibers, it results in aligned nanofibers. These aligned nanofibers offer specific advantages that make them suitable for various biomedical applications. The alignment mimics the natural structure of tissues, promoting cell adhesion, orientation, and migration. Fiber alignment can enhance the mechanical properties and offer controlled drug release.

Outlined in this review are some major biomedical applications of aligned nanofiber in the regeneration of bone, neuron, cardiac, and muscle tissue. Aligned nanofibers possess the mechanical properties and hierarchical structure of the natural bone. Aligned fibers provide guidance cues for the regeneration of neural tissues and can guide the growth of axons in a specific direction, which is essential for nerve regeneration. The use of aligned nanofibers in cardiac tissue engineering shows great promise. They replicate the anisotropic structure of the native myocardium, guiding the orientation of cardiac cells and promoting functional tissue regeneration. Incorporating conductive materials enhances electrical conductivity, which is important for supporting electrical signaling in cardiac tissue. Additionally, incorporating elements that promote angiogenesis and stimulate the vascular network can have a positive effect. The anisotropic structure and mechanical strength of aligned nanofibers facilitate the growth of muscle cells. They promote the alignment of myoblasts, which in turn supports the formation of aligned myotubes and muscle fibers.

However, there are certain limitations associated with the use of aligned nanofibers that must be taken into account. Reproducibility and scalability are some of the challenges that need to be considered while developing aligned nanofiber-based biomedical applications. Moreover, the choice of materials for electrospinning should be assessed carefully for their biocompatibility and degradation properties. Vascularization is crucial in tissue regeneration; different strategies must be employed to facilitate efficient nutrient and oxygen supply. Suitable porosity and interconnectivity are other critical factors for biomedical applications as they play a role in cellular filtration, proliferation, and migration. Tissue integration of fiber scaffold with the host tissue is also crucial for tissue functionality, which needs to be achieved. Preclinical studies and rigorous in vivo testing are necessary to assess the safety and efficacy of aligned nanofiber scaffolds for biomedical applications.